\documentclass[12pt,aps,prd,showpacs,amsmath,amssymb]{revtex4}
\input epsf
\usepackage{graphicx}
\allowdisplaybreaks

\begin{document}

\title{QCD sum rules study of $X(4350)$
\footnote{Supported by the National Natural Science
Foundation of China under Grant Nos. (11347174, 11275268, 11105222 and 11025242).}}
\author{Zeng Mo$^{\P}$, Chun-Yu Cui$^{\P}$\footnote{E-mail address: cycui@nudt.edu.cn}, Yong-Lu Liu$^*$ and Ming-Qiu Huang$^*$}
\affiliation{$^{\P}$ Department of Physics, School of Biomedical Engineering, Third Military Medical University, Chongqing 400038, China\\
$^*$ College of Science, National University of Defense Technology, Hunan 410073, China}

\begin{abstract}
The QCD sum rule approach is used to analyze the nature of the
rencently observed new resonance $X(4350)$, which is assumed to be a
diquark-antidiquark state $[cs][\bar{c}\bar{s}]$ with $J^{PC}=1^{-+}$. The
interpolating current representing this state is
proposed. In the calculation, contributions of operators up to
dimension six are included in the operator product expansion (OPE), as well as terms which are linear in the strange quark mass
$m_s$. We find $m_{1^{-+}}=(4.82\pm 0.19)\,\mbox{GeV}$, which is not compatible with the $X(4350)$ structure
as a $1^{-+}$ tetraquark state. Finally, we also discuss the difference of  a four-quark state's mass whether the state's interpolating current has a definite charge conjugation.
\end{abstract}
\maketitle
\section{Introduction}\label{sec1}
Recently, the Belle Collaboration found a new narrow structure $X(4350)$ in
the $\phi J/\psi$ mass spectrum, when searching for
$Y(4140)$ reported by CDF Collaboration. The mass and width of the
state is $(4350.6^{+4.6}_{-5.1}\pm 0.7)\,\rm{MeV}$ and
$(13.3^{+17.9}_{-9.1}\pm 4.1)\,\rm{MeV}$ \cite{bellex}. Some theoretical works have been done before the Belle
experiment. Wang performs a systematic study of the mass spectrum of the vector
hidden charm and bottom tetraquark states using the QCD sum rules~\cite{wang2}. Zhang $et$ $al.$ calculate the mass of the $D_{s}^{\ast}
\bar{D}_{s}^{\ast}$
molecular state to be $(4.36 \pm 0.08)\,\rm{GeV}$~\cite{Zhang}. Stancu studies mass spectrum
of the $[cs][\bar{c}\bar{s}]$ tetraquarks \cite{Stancu4}. Both of
their results are consistent with the experimental data \cite{bellex}. The possible quantum numbers for a state decaying
into $\phi J/\psi$ are $J^{PC}=0^{++},~1^{-+}$ or $2^{++}$. Wang interprets the $X(4350)$ as a scalar
$\bar{c}c$ and ${D}_s^\ast {\bar {D}}_s^\ast$ mixing state with
QCD sum rules \cite{wang}. In Ref.~\cite{Liu0911}, Liu $et$ $al.$ discuss the possibility that the $X(4350)$ is an excited $P$-wave charmonium state $\chi_{c2}''$  by studying the strong decays of the $P$-wave charmonium states with the $^3P_0$ model.

Among these quantum numbers, $J^{PC}=1^{-+}$ known as exotic attracts great theoretical attention. The state considered in Ref.~\cite{Zhang} has $J^P=1^-$ without a definite
charge conjugation. Using QCD sum rules~\cite{Nielsen}, Albuquerque $et$ $al.$ study a molecular state with a vector and a scalar $D_s$ mesons
with a definite positive charge conjugation, and conclude
that it is not possible to describe the $X(4350)$ structure as a
$1^{-+}~D_s^*{D}_{s0}^*$ molecular state. Otherwise, Ma uses effective
lagrangian approach to estimate $X(4350)$ decay, and concludes that
$X(4350)$ as a $D_s^*{D}_{s0}^*$ can't be ruled out \cite{Ma}. Under
such a circumstance, there is no definite structure with $J^{PC}=1^{-+}$ for $X(4350)$, we propose to take the $X(4350)$ as a diquark-antidiquark state with
$J^{PC}=1^{-+}$. Mass property is helpful for understanding whether $X(4350)$ could be a diquark-antidiquark state with
$J^{PC}=1^{-+}$. However, in low energy and hadronic scales, it is difficult to get reliable theoretical estimate for the mass using the perturbative QCD. Therefore, we need some non-perturative methods to describe the non-perturative phenomena. QCD sum rules \cite{svz,reinders,overview3} is powerful since they are based on the fundamental QCD lagrangian. From this perspective of view, the report \cite{NielsenPR} guides practitioners to compute masses of this kind of new discovering states.

The paper is organized as follows. In Sec. \ref{sec2}, QCD sum
rule for the diquark-antidiquark tetraquark state with $J^{PC}=1^{-+}$ state is introduced,
and both the phenomenological representation and QCD side are
derived. In Sec.\ref{sec3}, we present numerical analysis to extract the hadronic mass and decay constant. This section also contains a brief summary.

\section{the tetraquark state QCD sum rules}\label{sec2}
The lowest-dimension current interpolating a $J^{PC}=1^{-+}$ state with the symmetric
spin distribution $[cs]_{S=0}[\bar{c}\bar{s}]_{S=1}+
[cs]_{S=1}[\bar{c}\bar{s}]_{S=0}$ is given by
\begin{eqnarray}
j_\mu={\epsilon_{abc}\epsilon_{dec}\over\sqrt{2}}[(s_a^T C\gamma_5
c_b) (\bar{s}_d\gamma_\mu\gamma_5 C\bar{c}_e^T)-(s_a^T C \gamma_5
\gamma_\mu c_b) (\bar{s}_d \gamma_5 C \bar{c}_e^T)]\;, \label{field}
\end{eqnarray}
where $a,~b,~c,~...$ are color indices and $C$ is the charge
conjugation matrix.

The QCD sum rule attempts to link the hadron phenomenology with the
interactions of quarks and gluons, which contains three main
ingredients: an approximate description of the correlator in terms
of intermediate states through the dispersion relation, a
description of the same correlator in terms of QCD degrees of
freedom via an OPE, and a procedure for matching these two
descriptions and extracting the parameters that characterize the
hadronic state of interest.

The two-point correlation function is given by
\begin{eqnarray}
\Pi^{\mu\nu}(q^{2})=i\int
d^{4}x\mbox{e}^{iq.x}\langle0|T[j^{\mu}(x)j^{\nu+}(0)]|0\rangle.
\end{eqnarray}

Lorentz covariance implies that the two-point correlation function
can be generally parameterized as
\begin{eqnarray}
\Pi^{\mu\nu}(q^{2})=(\frac{q^{\mu}q^{\nu}}{q^{2}}-g^{\mu\nu})\Pi^{(1)}(q^{2})+\frac{q^{\mu}q^{\nu}}{q^{2}}\Pi^{(0)}(q^{2}).
\end{eqnarray}

The part of the correlator proportional to $g_{\mu\nu}$ will be
chosen to extract the mass sum rule, since it
gets contributions only from the $1^{-+}$ state. The coupling of the current with the state can be defined by the decay constant as follows:
\begin{eqnarray}
\langle 0|j_{\mu}|X\rangle&=&\lambda \epsilon_{\mu}.
\end{eqnarray}
In phenomenology, $\Pi^{(1)}(q^{2})$ can be expressed as a dispersion integral over a
physical spectral function
\begin{eqnarray}
\Pi^{(1)}(q^2)=\frac{\lambda^{2}}{M_{X}^{2}-q^{2}}+\frac{1}{\pi}\int_{s_{0}}
^{\infty}ds\frac{\mbox{Im}\Pi^{(1)\mbox{phen}}(s)}{s-q^{2}}+\mbox{subtractions},
\end{eqnarray}
where $M_{H}$ denotes the mass of the hadronic resonance. In the OPE
side, $\Pi^{(1)}(q^{2})$ can be written in terms of a dispersion
relation as
\begin{eqnarray}
\Pi^{(1)}(q^{2})=\int_{(2m_{c}+2m_{s})^{2}}^{\infty}ds\frac{\rho^{\mbox{\scriptsize
OPE}}(s)}{(s-q^{2})},
\end{eqnarray}
where the spectral density is given by
\begin{eqnarray}
\rho^{\mbox{\scriptsize
OPE}}(s)=\frac{1}{\pi}\mbox{Im}\Pi^{\mbox{(1)}}(s).
\end{eqnarray}

After equating the two sides, assuming quark-hadron duality, and making a Borel transformation, the sum rule can be written as
\begin{eqnarray}
\lambda^{2}e^{-M_{X}^{2}/M^{2}}&=&\int_{(2m_{c}+2m_{s})^{2}}^{s_{0}}ds\rho^{\mbox{\scriptsize
OPE}}(s)e^{-s/M^{2}}.
\end{eqnarray}

To eliminate the decay constant $\lambda$, one
reckons the ratio of the derivative of the sum rule and itself, and then
yields
\begin{eqnarray}\label{sum rule 1}
M_{X}^{2}&=&\int_{(2m_{c}+2m_{s})^{2}}^{s_{0}}ds\rho^{\mbox{\scriptsize
OPE}}s e^{-s/M^{2}}/
\int_{(2m_{c}+2m_{s})^{2}}^{s_{0}}ds\rho^{\mbox{\scriptsize
OPE}}e^{-s/M^{2}}.
\end{eqnarray}

When calculating the OPE side, we work at the leading order in $\alpha_{s}$
and consider condensates up to dimension six, with the similar
techniques in Refs.~\cite{technique,technique1}. The $s$ quark is
regarded as a light one and the terms are considered up to the order of
$m_{s}$. To keep the heavy-quark mass finite, one uses the
momentum-space expression for the heavy-quark propagator. One
calculates the light-quark part of the correlation function in the
coordinate space, which is then Fourier-transformed to the momentum
space in $D$ dimension. The resulting light-quark part is combined
with the heavy-quark part before it is dimensionally regularized at
$D=4$. For the heavy-quark propagator with two and three gluons
attached, the momentum-space expressions given in Ref.~\cite{reinders} are used.
After some tedious OPE calculations, the concrete forms of spectral
densities can be derived.

\begin{eqnarray}
\rho^{\mbox{\scriptsize
OPE}}(s)=\rho^{\mbox{pert}}(s)+\rho^{\langle\bar{s}s\rangle}(s)+\rho^{\langle
g^{2}G^{2}\rangle}(s)+\rho^{\langle
g\bar{s}\sigma\cdot G s\rangle}(s)+\rho^{\langle\bar{s}s\rangle^{2}}(s),\nonumber
\end{eqnarray}

\begin{eqnarray}
\rho^{\mbox{pert}}(s)&=&\frac{1}{3*2^{10}\pi^{6}}\int_{\alpha_{min}}^{\alpha_{max}}\frac{d\alpha}{\alpha^{3}}\int_{\beta_{min}}^{1-\alpha}\frac{d\beta}{\beta^{3}}(1-\alpha-\beta)[(\alpha+\beta)m_{c}^{2}-\alpha\beta
s]^{3}\nonumber\\&&{}
[(5\alpha^{2}+10\alpha\beta-\alpha+5\beta^{2}-\beta+2)m_{c}^{2}-3\alpha\beta s(\alpha+\beta+1)]\nonumber\\&&{}
+\frac{1}{2^{8}\pi^{6}}m_{c}m_{s}\int_{\alpha_{min}}^{\alpha_{max}}\frac{d\alpha}{\alpha^{3}}\int_{\beta_{min}}^{1-\alpha}\frac{d\beta}{\beta^{2}}(1-\alpha-\beta)^2(\alpha+\beta)[(\alpha+\beta)m_{c}^{2}-\alpha\beta
s]^{3},\nonumber
\end{eqnarray}

\begin{eqnarray}
\rho^{\langle\bar{s}s\rangle}(s)&=&\frac{\langle\bar{s}s\rangle}{2^{5}\pi^{4}}m_{c}^{2}m_{s}\int_{\alpha_{min}}^{\alpha_{max}}\frac{d\alpha}{\alpha}\int_{\beta_{min}}^{1-\alpha}\frac{d\beta}{\beta}(-3-\alpha-\beta)[(\alpha+\beta)m_{c}^{2}-\alpha\beta
s]\nonumber\\&&{}
+\frac{\langle\bar{s}s\rangle}{2^{5}\pi^{4}}m_{c}\int_{\alpha_{min}}^{\alpha_{max}}\frac{d\alpha}{\alpha^{2}}\int_{\beta_{min}}^{1-\alpha}\frac{d\beta}{\beta^{2}}(1-\alpha-\beta)(\alpha+\beta)[(\alpha+\beta)m_{c}^{2}-\alpha\beta
s]^{2}\nonumber\\&&{}
+\frac{\langle\bar{s}s\rangle}{2^{5}\pi^{4}}m_{s}\int_{\alpha_{min}}^{\alpha_{max}}\frac{d\alpha}{\alpha(1-\alpha)}[m_{c}^{2}-\alpha(1-\alpha)
s]^{2},\nonumber
\end{eqnarray}

\begin{eqnarray}
\rho^{\langle g^{2}G^{2}\rangle}(s)&=&\frac{\langle
g^{2}G^{2}\rangle}{9*2^{10}\pi^{6}}m_{c}^{4}\int_{\alpha_{min}}^{\alpha_{max}}{d\alpha}\int_{\beta_{min}}^{1-\alpha}\frac{d\beta}{\beta^{3}}(1-\alpha-\beta)^3\nonumber\\&&{}
+\frac{\langle
g^{2}G^{2}\rangle}{3*2^{9}\pi^{6}}m_{c}^{3}m_{s}\int_{\alpha_{min}}^{\alpha_{max}}{d\alpha}\int_{\beta_{min}}^{1-\alpha}\frac{d\beta}{\beta^{2}}(1-\alpha-\beta)^2\nonumber\\&&{}
+\frac{\langle
g^{2}G^{2}\rangle}{3*2^{9}\pi^{6}}m_{c}^{3}m_{s}\int_{\alpha_{min}}^{\alpha_{max}}d\alpha\alpha\int_{\beta_{min}}^{1-\alpha}\frac{d\beta}{\beta^{3}}(1-\alpha-\beta)^2\nonumber\\&&{}
+\frac{\langle
g^{2}G^{2}\rangle}{3*2^{10}\pi^{6}}m_{c}^{2}\int_{\alpha_{min}}^{\alpha_{max}}\frac{d\alpha}{\alpha}\int_{\beta_{min}}^{1-\alpha}\frac{d\beta}{\beta^{3}}(1-\alpha-\beta)\nonumber\\&&{}
(3\alpha^{3}+4\alpha\beta+\beta^{2}-2\beta+1)[(\alpha+\beta)m_{c}^{2}-\alpha\beta
s]\nonumber\\&&{}
+\frac{\langle
g^{2}G^{2}\rangle}{2^{9}\pi^{6}}m_{c}m_{s}\int_{\alpha_{min}}^{\alpha_{max}}{d\alpha}\int_{\beta_{min}}^{1-\alpha}\frac{d\beta}{\beta^{3}}(1-\alpha-\beta)^2[(\alpha+\beta)m_{c}^{2}-\alpha\beta
s],\nonumber
\end{eqnarray}

\begin{align}
\rho^{\langle g\bar{s}\sigma\cdot G s\rangle}(s)&=\frac{\langle
g\bar{s}\sigma\cdot G
s\rangle}{3*2^{6}\pi^{4}}\int_{\alpha_{min}}^{\alpha_{max}}{d\alpha}
\int_{\beta_{min}}^{1-\alpha}{d\beta}m_{c}^2m_{s}
\nonumber\\ &{}+\frac{\langle g\bar{s}\sigma\cdot
Gs\rangle}{2^{6}\pi^{4}}m_{c}\int_{\alpha_{min}}^{\alpha_{max}}\frac{d\alpha}{\alpha}\int_{\beta_{min}}^{1-\alpha}\frac{d\beta}{\beta}(\alpha+\beta)[(\alpha+\beta)m_{c}^{2}-\alpha\beta
s]\nonumber\\ &{}+\frac{\langle g\bar{s}\sigma\cdot G
s\rangle}{3*2^{8}\pi^{4}}m_{s}\int_{\alpha_{min}}^{\alpha_{max}}{d\alpha}\int_{\beta_{min}}^{1-\alpha}\frac{d\beta}{\beta}[(\alpha+\beta)m_{c}^{2}-\alpha\beta
s]\nonumber\\&{} -\frac{\langle g\bar{s}\sigma\cdot G
s\rangle}{3*2^{3}\pi^{4}}m_{c}^2m_{s}\sqrt{1-4m_{c}^{2}/s}\nonumber\\&{}
+\frac{7\langle g\bar{s}\sigma\cdot G
s\rangle}{3*2^{6}\pi^{4}}m_{s}\int_{\alpha_{min}}^{\alpha_{max}}{d\alpha}[m_{c}^{2}-\alpha(1-\alpha)s]\nonumber\\&{}
-\frac{\langle g\bar{s}\sigma\cdot G
s\rangle}{3*2^{8}\pi^{4}}m_{s}\int_{\alpha_{min}}^{\alpha_{max}}\frac{d\alpha}{(1-\alpha)}[m_{c}^{2}-\alpha(1-\alpha)
s],\nonumber
\end{align}

\begin{eqnarray}
\rho^{\langle\bar{s}s\rangle^{2}}(s)&=&-\frac{\langle\bar{s}s\rangle^{2}}{3*2^{3}\pi^{2}}m_{c}^{2}\sqrt{1-4m_{c}^{2}/s}
+\frac{\langle\bar{s}s\rangle^{2}}{3*2^{3}\pi^{2}}m_{c}m_{s}\sqrt{1-4m_{c}^{2}/s}\nonumber\\&&{}
-\frac{\langle\bar{s}s\rangle^{2}}{3^{2}*2^{3}\pi^{2}}(\frac{s}{2}+m_{c}^{2})\sqrt{1-4m_{c}^{2}/s}.\nonumber
\end{eqnarray}

The integration limits are given
by $\alpha_{min}=(1-\sqrt{1-4m_{c}^{2}/s})/2$,
$\alpha_{max}=(1+\sqrt{1-4m_{c}^{2}/s})/2$, and $\beta_{min}=\alpha
m_{c}^{2}/(s\alpha-m_{c}^{2})$.

\section{Numerical analysis}\label{sec3}
This part is a numerical analysis of the sum rule (\ref{sum rule 1}). The
input values are taken as $m_{c}=1.23~\mbox{GeV}$, $m_{s}=0.13~\mbox{GeV}$,
$\langle\bar{q}q\rangle=-(0.23)^{3}~\mbox{GeV}^{3}$,
$\langle\bar{s}s\rangle=0.8~\langle\bar{q}q\rangle$, $\langle
g\bar{s}\sigma\cdot G s\rangle=m_{0}^{2}~\langle\bar{s}s\rangle$,
$m_{0}^{2}=0.8~\mbox{GeV}^{2}$, and $\langle
g^{2}G^{2}\rangle=0.88~\mbox{GeV}^{4}$ \cite{overview2,parameters}. Complying with the standard
procedure of the sum rule, the threshold $s_{0}$ and Borel
parameter $M^{2}$ are varied to find the optimal stability window,
in which the perturbative contribution should be larger than the
condensate contributions while the pole contribution should be larger than
continuum contribution. The continuum thresholds $s_0$ is not completely arbitrary as it is correlated to the
energy of the first excited state with the same quantum numbers as the state we considered, which is given by $s_0=(M_{H} + \Delta_{s})^2$. In many cases, the central value of $\sqrt{s_0}$ is connected to the mass $M_{H}$ of the studied state by the relation that the attained
mass value should be around $0.5\, \mbox{GeV}$ smaller than $\sqrt{s_0}$.
\begin{figure}
  \includegraphics[width=15cm]{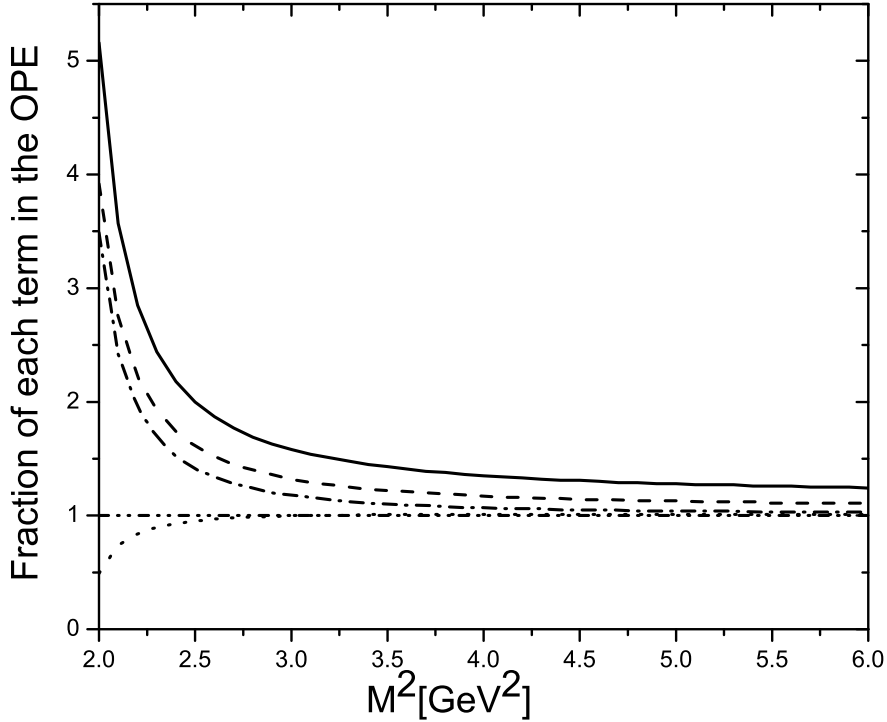}
 \caption{The relative contributions of different terms in the OPE for the $J^{PC}=1^{-+}$ tetraquark
state in the region
$1.5~\mbox{GeV}^2 \leq M^2\leq6.5\,\mbox{GeV}^2$ for $\sqrt{s_0} = 5.3$\mbox{GeV}.  We plot the
relative contributions starting with the perturbative contribution plus
$m_s$ correction (solid), and each other line represents the
relative contribution after adding of one extra condensate in the expansion:
+ $\langle\bar{s}s\rangle+m_{s}\langle\bar{s}s\rangle$  (dashed line),
+ $\langle
g^{2}G^{2}\rangle$ (dotted line), + $\langle
g\bar{s}\sigma\cdot G s\rangle+m_{s}\langle
g\bar{s}\sigma\cdot G s\rangle$
(dash-dotted line), + ${\langle\bar{s}s\rangle}^2+m_{s}{\langle\bar{s}s\rangle}^2$ (dash-dot-dotted line).}
\label{Figure1}
\end{figure}

\begin{figure}
  \includegraphics[width=15cm]{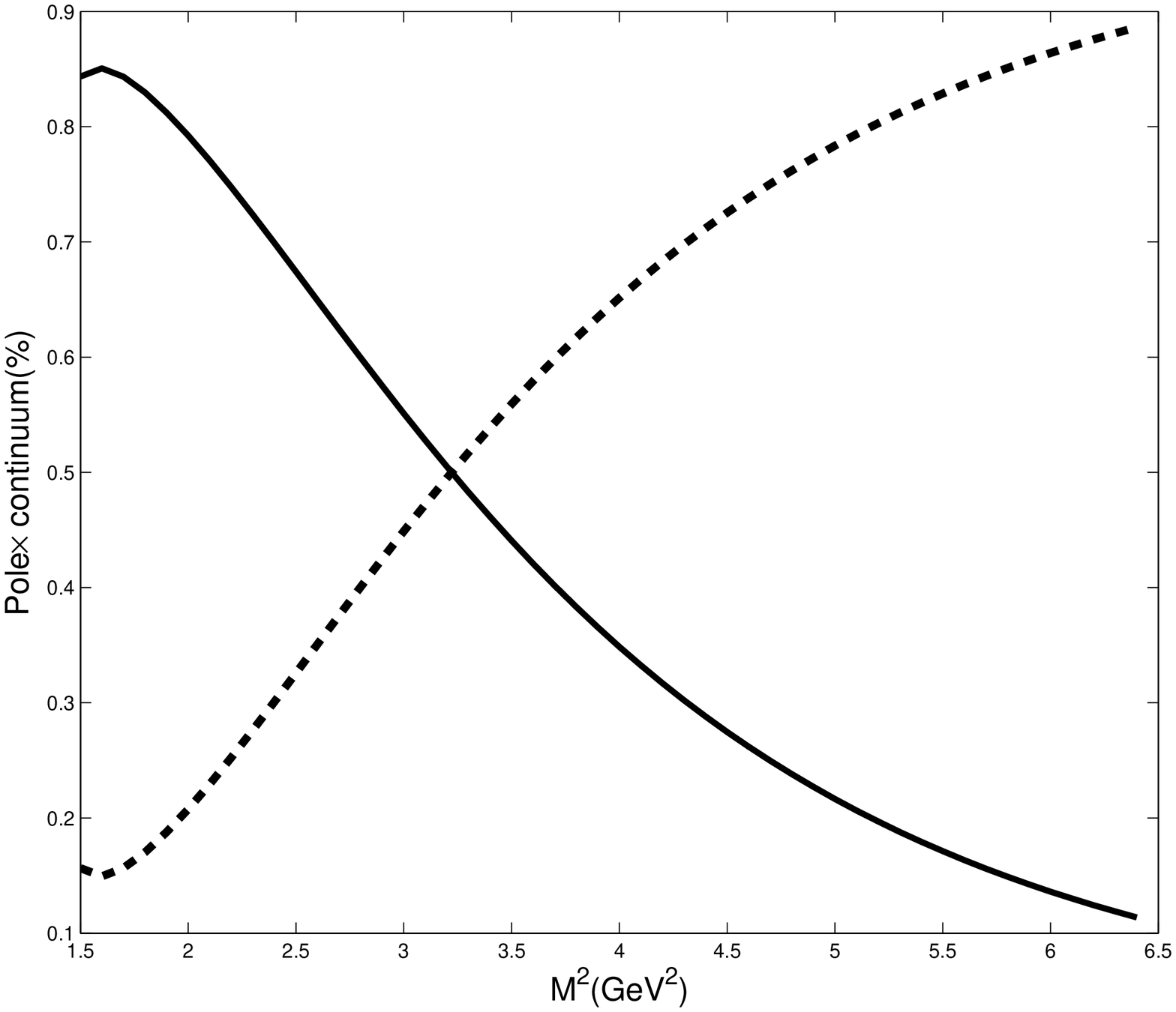}
 \caption{The solid line shows the relative pole contribution (the
pole contribution divided by the total contribution) and the dashed line shows the relative continuum
contribution for $\sqrt{s_0}=5.3~\mbox{GeV}$.}
\label{Figure2}
\end{figure}

We fix the central value at the point $\sqrt{s_0}=5.3~\mbox{GeV}$. In Fig.~\ref{Figure1} we plot contributions of all the terms in the
OPE side of the sum rule. From this figure it can be seen that for $M^2\geq 2.5\,
\mbox{GeV}^2$, the contribution of the dimension-6 condensate is less than 10\% of the
total contribution, which indicates a good Borel convergence. Therefore, we  fix the lower
value of $M^2$ in the sum rule window as $M^2_{min}= 2.5 $\mbox{GeV}$^2$.

\begin{figure}
\includegraphics[width=15cm]{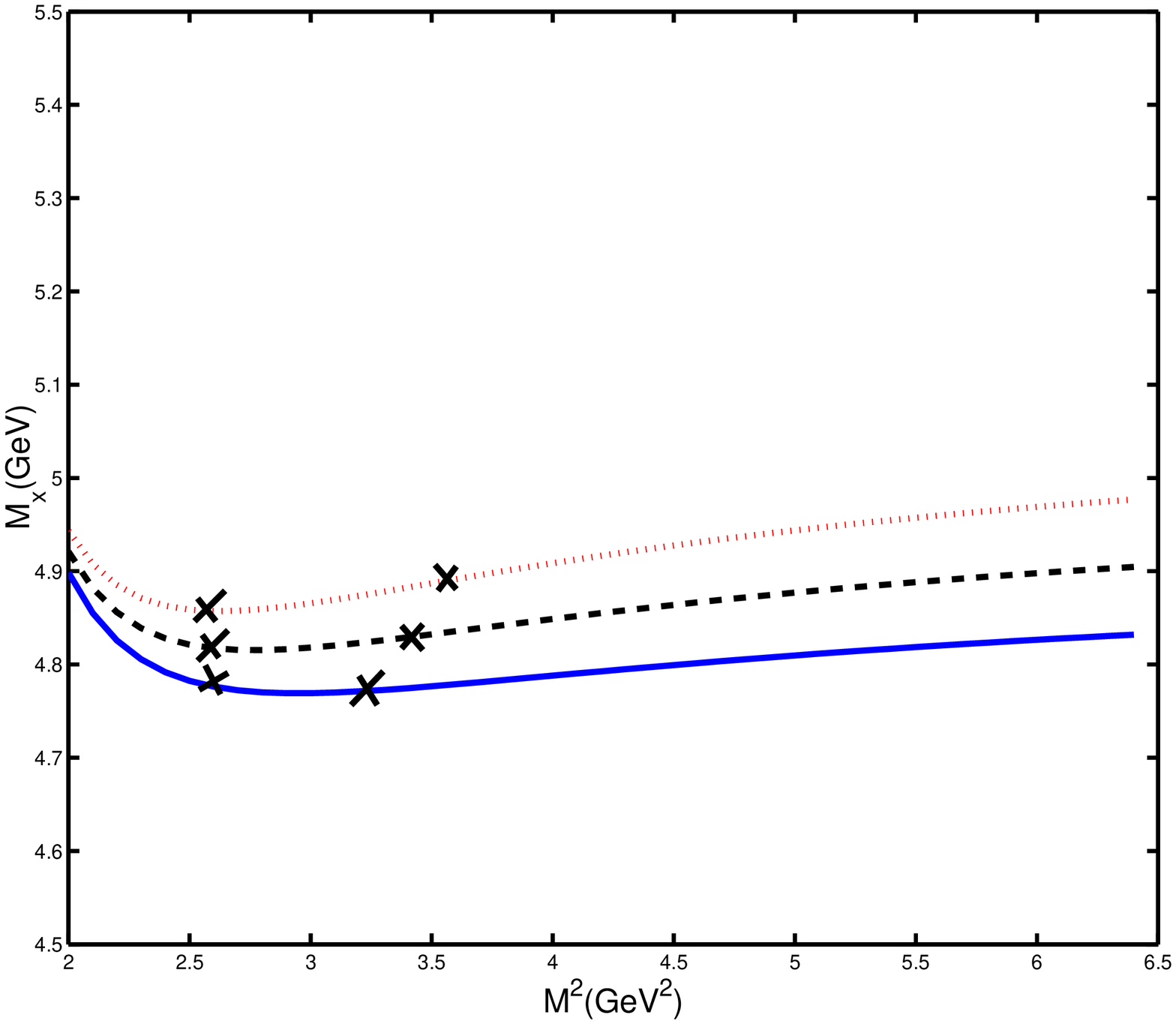}
\caption{The X state mass, described with a $J^{PC}=1^{-+} $ tetraquark
current, as a function of the sum rule parameter
($M^2$) for $\sqrt{s_0} =5.2$ GeV (solid line), $\sqrt{s_0} =5.3$ GeV (dotted
line), and $\sqrt{s_0} =5.4$ GeV (dashed line). The crosses
indicate the upper and lower limits in the Borel region.}
\label{Figure3}
\end{figure}

Fig.~\ref{Figure2} shows that the contributions from the pole terms and continuum terms
with variation of the Borel parameter $M^2$. The pole
contribution is bigger than the continuum contribution for $M^2\leq3.3\,\mbox{GeV}^2$. Therefore, we fix $M^2_{max}= 3.3\,\mbox{GeV}^2$ as the upper limit of the Borel window for $\sqrt{s_0}=5.3~\mbox{GeV}$. With the same analysis for the continuum threshold $\sqrt{s_0}=(5.3\pm0.1)\mbox{GeV}$, we determine the corresponding Borel windows in Fig.~\ref{Figure3}. In this figure, we show the mass of X state as functions of the Borel mass for  several threshold values $\sqrt{s_0}$.
It can be seen that we get a very good Borel stability for $M_x$. It is worth noting that errors in our results merely come from  the threshold
$s_{0}$ and the Borel parameter $M^{2}$, without involving the ones from
the variation of quark masses and QCD input parameters.

\begin{figure}
\includegraphics[width=15cm]{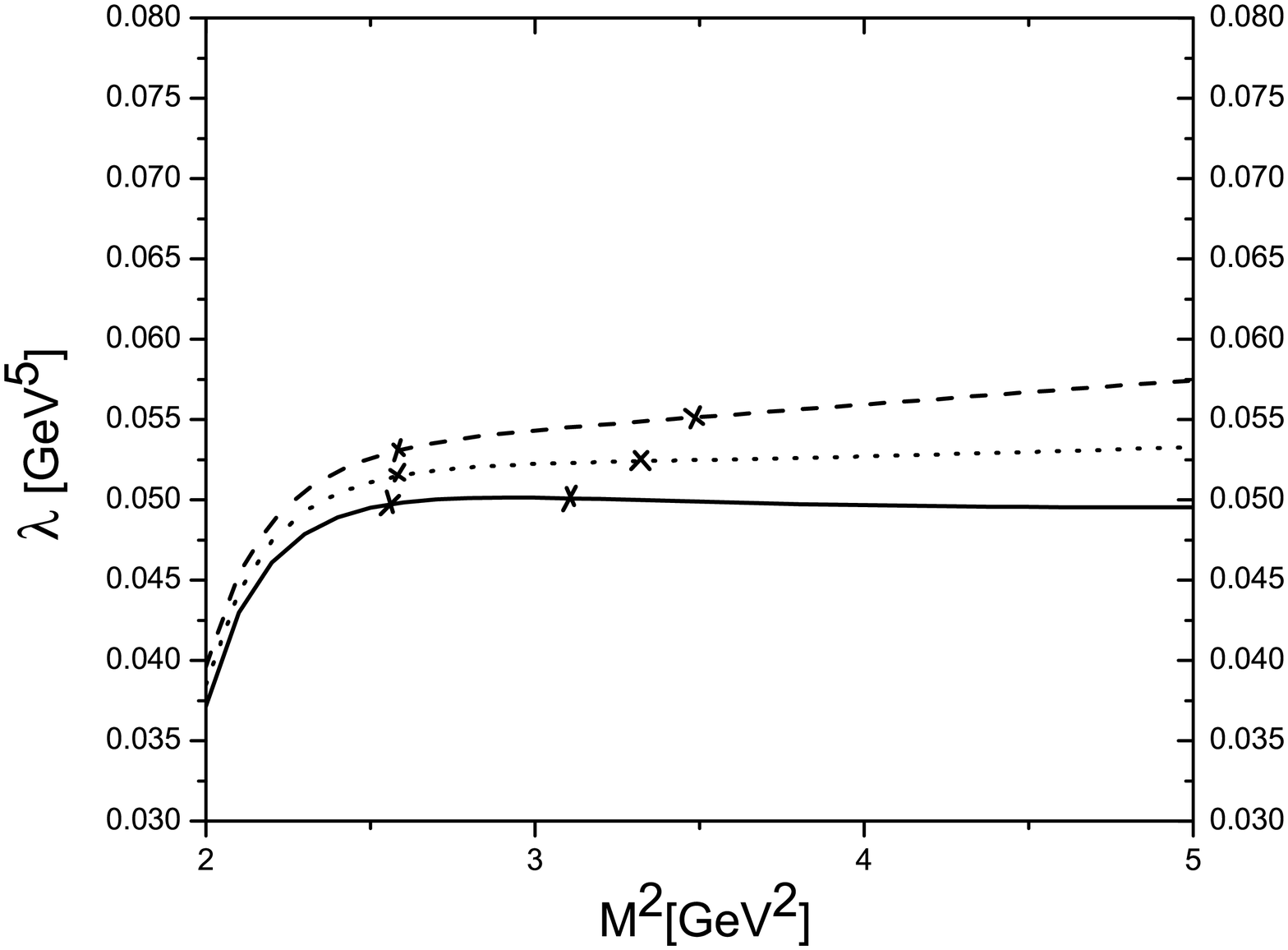}
\caption{The X state's decay constant as a function of the sum rule parameter
($M^2$) for $\sqrt{s_0} =5.2$ GeV (solid line), $\sqrt{s_0} =5.3$ GeV (dotted
line), and $\sqrt{s_0} =5.4$ GeV (dashed line). The crosses
indicate the upper and lower limits in the Borel region.}
\label{Figure4}
\end{figure}
Taking into account the uncertainties given above, we obtain the mass and decay constant of $X$
\begin{eqnarray}
M_{x}&=&(4.82\pm 0.19)\,\rm{GeV} \, , \nonumber\\
\lambda&=&(5.23 \pm 0.28)\times 10^{-2}\,\rm{GeV^5} \, .
\label{decaycons}
\end{eqnarray}

The mass obtained is not compatible with the mass of the narrow structure $X(4350)$
observed by Belle. It is, however, very interesting to notice that the mass
obtained for a diquark-antidiquark state $c \bar{c} s \bar{s}$ with $J^{PC}=1^{--}$ is $m_{1^{--}}=(4.65\pm0.10)~\mbox{GeV}$~\cite{Nielsen1}. It is  smaller than what we have obtained with the $J^{PC}=1^{-+} $ tetraquark current. This may be an indication that it is easier to form tetraquark states with non-exotic quantum numbers. It is just as what the authors found in Ref.~\cite{Nielsen} that the mass of the molecular sate $D_s^{*+}\bar{D}_{s0}^{-}$ with $J^{PC}=1^{--}$ is lower than that with $J^{PC}=1^{-+}$. It is a hint that it is easier to form molecular sate states with non-exotic quantum numbers. We also notice that, Wang \cite{wang1} study $J^{P}=1^{-}$ tetraquark state whose interpolating current doesn't have a definite charge conjugation using QCD sum rules, and obtained $m_{1^{-}}=(5.16\pm 0.16)\,\mbox{GeV}$. It is much larger than the results with a definite charge conjugation. Opposite to the $J^{P}=1^{-}$ $D_{s}^*D_{s0}^*$ molecular state circumstance, wherein, the state~\cite{Zhang} without a definite charge conjugation interpolator is much smaller than states with a definite charge conjugation interpolator~\cite{Nielsen}.

In summary, by assuming $X(4350)$ as a $[cs][\bar{c}\bar{s}]$ tetraquark state with quantum numbers $J^{PC}=1^{-+}$, the QCDSR approach has been applied to calculate the mass of the resonance.
Our numerical result is $m_{X}=(4.82\pm0.19)~\mbox{GeV}$, which disfavors the $X(4350)$ observed by the Belle is a $J^{PC}=1^{-+}$ tetraquark state.

\begin{acknowledgments}
This work was supported in part by the National Natural Science
Foundation of China under Contract Nos.11347174, 11275268, 11105222 and 11025242.
\end{acknowledgments}


\begin{thebibliography}{99}
\bibitem{bellex}C.~P.~Shen {\it et al.}, (Belle Collaboration), Phys.
Rev. Lett. {\bf104}, 112004 (2010).

\bibitem{wang2}Z.-G. Wang, J. Phys. G {\bf36}, 085002 (2009).

\bibitem{Zhang} J. R. Zhang and M. Q. Huang, Commun. Theor. Phys. {\bf54}, 1075 (2010)

\bibitem{Stancu4} F. Stancu, J. Phys. G. {\bf37}, 075017 (2010)

\bibitem{wang} Z.-G. Wang, Phys. Lett. B {\bf690}, 403 (2010).

\bibitem{Liu0911} X. Liu, Z. G. Luo and Z. F. Sun, Phys. Rev. Lett. {\bf 104}, 122001 (2010).

\bibitem{Nielsen}R.M. Albuquerque, J.Dias and M. Nielsen, Phys. Lett. B {\bf690}, 141 (2010).

\bibitem{Ma} Y.-L. Ma, Phys. Rev. D {\bf82}, 015013 (2010)

\bibitem{svz} M.~A.~Shifman, A.~I.~Vainshtein, and V.~I.~Zakharov, Nucl. Phys. {\bf B147}, 385 (1979); {\bf B147}, 448 (1979);
 V.~A.~Novikov, M.~A.~Shifman, A.~I.~Vainshtein, and V.~I.~Zakharov, Fortschr. Phys. {\bf 32}, 585 (1984), M.~A.~Shifman, Vacuum Structure and QCD Sum Rules, North-Holland, Amsterdam 1992.
\bibitem{reinders}L.~J.~Reinders, H.~R.~Rubinstein, and S.~Yazaki, Phys. Rep. {\bf 127}, 1 (1985).
\bibitem{overview3}P.~Colangelo and A.~Khodjamirian, in: M.~Shifman (Ed.), At the Frontier of Particle
Physics: Handbook of QCD, vol. 3, Boris Ioffe Festschrift, World
Scientific, Sigapore, 2001, pp. 1495-1576, arXiv:0010175;
A.~Khodjamirian, talk given at Continuous Advances in QCD
2002/ARKADYFEST, arXiv:0209166.

\bibitem{NielsenPR}M.~Nielsen, F.~S.~Navarra, and S.H. Lee,
Phys. Rep. {\bf 497}, 41 (2010).

\bibitem{technique}R.~D.~Matheus, S.~Narison, M.~Nielsen, and J.~M.~Richard, Phys. Rev. D {\bf75}, 014005 (2007);
 S.~H.~Lee, K.~Morita, and M.~Nielsen, Phys.
Rev. D {\bf78}, 076001 (2008); M.~E.~Bracco, S.~H.~Lee, M.~Nielsen,
R.~R.~daSilva, Phys. Lett. B {\bf671}, 240 (2009); S.~H.~Lee,
K.~Morita, and M.~Nielsen, Nucl. Phys. A {\bf815}, 29 (2009);
R.~M.~Albuquerque and M.~Nielsen, Nucl. Phys. A {\bf815}, 53 (2009).

\bibitem{technique1}C. Y. Cui, Y. L. Liu, M. Q. Huang, Phys. Rev. D 85, 074014 (2012); C. Y. Cui, Y. L. Liu, G. B. Zhang, M. Q. Huang, Commun. Theor. Phys. 57 (2012) 1033-1036; C. Y. Cui, Y. L. Liu, W. B. Chen, M. Q. Huang, arXiv:1304.1850 [hep-ph]; C. Y. Cui, Y. L. Liu, M. Q. Huang, arXiv:1308.3625 [hep-ph]; J.~R.~Zhang and M.~Q.~Huang, Phys. Rev. D {\bf 77}, 094002 (2008); Phys. Rev. D {\bf 78}, 094007 (2008); Phys. Lett. B {\bf 674}, 28 (2009); J. Phys. G {\bf37}, 025005 (2010).

\bibitem{overview2}S.~Narison, QCD Spectral Sum Rules, World Scientific,Singapore,1989.

\bibitem{parameters}S.~Narison, Monogr. Part. Phys. Nucl. Phys. Cosmol. {\bf 17} 1 (2002).

\bibitem{Nielsen1}R.~M.~Albuquerque and M.~Nielsen, Nucl. Phys. A {\bf 815}, 53 (2009).

\bibitem{wang1}Z.-G. Wang, Eur. Phys. J. C {\bf59}, 675 (2009).
\end{thebibliography}
\end{document}